\documentclass[12pt]{article}

\usepackage{amsmath,cite}
\usepackage{amssymb}
\usepackage[hidelinks]{hyperref}
\usepackage{cite}
\usepackage[latin1]{inputenc} % Darstellung von Umlauten
\usepackage{color}
\usepackage[english	]{babel}
\usepackage{array}
\usepackage{float}
\usepackage[pdftex]{graphicx}
\usepackage{mathtools}
\usepackage{physics}
\usepackage{stmaryrd}
\usepackage{amssymb}
\usepackage{graphics}
\usepackage{tikz}
\usepackage{geometry}
\usepackage{bbold}
\newcommand{\p}[1]{(\ref{#1})}
\newcommand{\be}{\begin{equation}}
\newcommand{\ee}{\end{equation}}
\newcommand{\bea}{\begin{eqnarray}}
\newcommand{\eea}{\end{eqnarray}}

\newcommand{\besubeqs}{\begin{subequations}}
\newcommand{\esubeqs}{\end{subequations}}

\begin{document}
\setcounter{page}0
\renewcommand{\thefootnote}{\fnsymbol{footnote}} 
\begin{titlepage}
\vskip .7in
\begin{center}
{\Large \bf  A Note on Shape Invariant Potentials for
Discretized Hamiltonians
 } \vskip .7in 
 {\Large 
Jonas Sonnenschein$^1$\footnote{e-mail: {\tt  jonas.sonnenschein1@oist.jp }} and  
Mirian Tsulaia$^2$\footnote{e-mail: {\tt  mirian.tsulaia@oist.jp }} 
}
\vskip .4in {  $^1$\it Theory of Quantum Matter Unit, Okinawa Institute of Science and Technology, \\ 1919-1 Tancha, Onna-son, Okinawa 904-0495, Japan\\
$^2$\it Quantum Gravity Unit, Okinawa Institute of Science and Technology, \\ 1919-1 Tancha, Onna-son, Okinawa 904-0495, Japan}\\
\vskip .8in
\begin{abstract}

Using the method of the  "exact discretization"
of the Schr\"odinger equation,
 we propose  a particular discretized version of the $N=2$
Supersymmetric Quantum Mechanics.
After defining the corresponding  shape invariance condition,
we show that the energy spectra and wavefunctions 
for discretized Quantum Mechanical systems
can be found using the technique of $N=2$ Supersymmetric Quantum Mechanics
exactly the same way as it is done for their  continuous counterparts.
As a demonstration of the present method, we find the energy spectrum
for a discretized Coulomb potential and its ground state wave function.
\end{abstract}

\end{center}

\vfill

\end{titlepage}

\renewcommand{\thefootnote}{\arabic{footnote}}
\setcounter{footnote}0

\section{Introduction}

 $N=2$ Supersymmetric Quantum Mechanics (SQM),
 \cite{Witten:1981nf} --\cite{Witten:1982im}
 being originally introduced as a tool for studying
 of the problem of supersymmetry breaking
 in more complicated supersymmetric field theories,
 has many interesting applications both
 in Physics and Mathematics (see for example
 \cite{Cooper:1994eh} -- \cite{deLimaRodrigues:2002ya} for  reviews).
 
 One such application is finding of the energy spectra
 of ``usual" nonsupersymmetric Hamiltonians by using supersymmetry as an
 auxiliary tool. Recall 
 that $N=2$ supersymmetry algebra contains  mutually  
 hermitian conjugated supercharges $Q$ and $ Q^\dagger$
and the Hamiltonian $H$,
with only nonzero anticommutation relation
 \be \label{alg}
 \{ Q,  Q^\dagger \} = H.
 \ee
 This algebra can be realized in a matrix form as follows:
  \begin{align}
Q = \begin{pmatrix}
0 & 0\\
A_1 & 0
\end{pmatrix}, \quad Q^\dagger = \begin{pmatrix}
0 & A^\dagger_1\\
0 & 0
\end{pmatrix},
\quad H = \begin{pmatrix}
H^{(1)}-E_0^{(1)} & 0 \\
0 & H^{(2)}-E_0^{(1)}
\end{pmatrix}.
\end{align}
 In fact, the supersymmetric Hamiltonian $H$
 splits into two 
 Hamiltonians $H^{(1)}$ and $H^{(2)}$, each being factorized 
 in terms of a ground state energy $E_0^{(1)}$ for the first Hamiltonian
  and
 two ladder operators $A_1^\dagger$ and $ A_1$
 \be \label{fact}
 H^{(1)} = A_1^\dagger A_1 + E_0^{(1)}, \quad H^{(2)} = A_1 A_1^\dagger
 + E_0^{(1)}.
 \ee
  For the case of one dimensional space, the operators $A_1^\dagger$ and $A_1$ 
 have the form:
 \be \label{A12}
 A_1^\dagger= \frac{\hbar}{\sqrt{2m}}\frac{d}{d x} + W_1(x), \quad A_1= - \frac{\hbar}{\sqrt{2m}}\frac{d}{d x} + W_1(x),
 \ee
 where an arbitrary function $W_1(x)$, called the superpotential,
 is expressed via the ground state wavefunction of the first Hamiltonian
 \be \label{sp}
 W_1(x) = \frac{\hbar}{\sqrt{2m}} \frac{d \ln \psi_0^{(1)}(x)}{d x}. 
 \ee
 As a consequence of relations \p{fact}, 
 the Hamiltonians $H^{(1)}$ and $H^{(2)}$ have the following form:
 \be \label{ph}
 H^{(1,2)} = -\frac{d^2}{d x^2} + V^{(1,2)}(x),
 \ee
 where
 \bea \label{potentials} 
 &&V^{(1)}(x)= W_1^2(x) - \frac{\hbar}{\sqrt{2m}}\frac{d W_1(x)}{d x} + E_0^{(1)}, \\ \nonumber
 &&V^{(2)}(x)= W_1^2(x) + \frac{\hbar}{\sqrt{2m}} \frac{d W_1(x)}{d x} + E_0^{(1)}.
 \eea
 As it can be concluded from the discussion above,  $N=2$ SQM
 has the following basic properties:
 \begin{itemize}
 
 \item From the algebra \p{alg} it follows that
 the spectrum of the Hamiltonian $H$ is positive semi-definite.
 
 \item The wave functions and the corresponding energy levels of the 
 Hamiltonians \p{ph} are related as 
 \be
   \psi_{n}^{(2)} = (E_{n+1}^{(1)}-E_{0}^{(1)}) A_1 \psi_{n+1}^{(1)},
   \quad  E_{n}^{(2)} = E_{n+1}^{(1)}.
 \ee
 \item If one of the Hamiltonians (say $H^{(1)}$) has an eigenstate with zero energy,
 then the corresponding wave function can be found from the equation
 \be \label{exactsusy}
 A_1 \psi_{0}^{(1)} =0
 \ee
  After using the explicit form of $A_1$ given in
 \p{A12}, this equation leads to \p{sp}.
 \item The Hamiltonian $H^{(2)}$ can be ``refactorized"
 \be
  H^{(2)} = A_2^\dagger A_2 + E_0^{(2)} = A_2^\dagger A_2 + E_1^{(1)}
 \ee
in terms of  
new ladder operators
\bea \label{a212}
&&A_2^\dagger = \frac{\hbar}{\sqrt{2m}} \left (\frac{d}{d x}  + \frac{d \ln \psi_0^{(2)}(x)}{d x}
\right ), \\ \nonumber 
 &&A_2  = \frac{\hbar}{\sqrt{2m}} \left (-\frac{d}{d x}  + \frac{d \ln \psi_0^{(2)}(x)}{d x}
\right )
 \eea
and a new  superpotential, which is expressed  in terms of the lowest
energy wavefunction of the Hamiltonian $H^{(2)}$.
 Then one can construct the third Hamiltonian $H^{(3)}$
 \be
  H^{(3)} = A_2 A_2^\dagger  + E_0^{(2)} = A_2 A_2^\dagger  + E_1^{(1)},
 \ee
 which has the same energy levels as $H^{(2)}$
 except of the lowest energy of $H^{(2)}$.
 Continuing in this manner
 one can construct a hierarchy of  Hamiltonians.
 The total number of the Hamiltonians in the hierarchy
 is equal to 
  a number $n$ of the bound states
 of the Hamiltonian $H^{(1)}$. The energy levels and the wavefunctions  of these Hamiltonians being 
 \bea \label{en-lev-cont}
 &&E_n^{(m)}= E_{n+1}^{(m-1)}=...=E_{n+m-1}^{(1)}, \\ \nonumber
 &&\psi_n^{(m)} = (E_{n+m-1}^{(1)}-E_{m-2}^{(1)})^{-1/2}
... (E_{n+m-1}^{(1)}-E_{0}^{(1)})^{-1/2}
 A_{m-1}...A_1 \, \psi^{(1)}_{n+m-1}.
 \eea
\item Suppose the potentials in \p{potentials} satisfy the so called shape invariance condition
\cite{Gendenshtein:1983skv}
\be \label{SI-cont}
V^{(2)} (x, a_1)= V^{(1)} (x, a_2) + R(a_1),
\ee
where the parameters $a_1$ and $a_2$
are related to each other via some function $a_2= \phi (a_1)$
and the ``rest" $R(a_1)$ does not depend on the coordinate $x$. 
Then the spectrum of the Hamiltonian $H^{(1)}$ can be found
to be
\be \label{cont-sp}
E_n^{(1)} = \sum_{k=1}^n R(a_k).
\ee
Here,  $a_k$ means that the function $\phi(a_1)$ is applied $k$
times.
To obtain this result, one  constructs a hierarchy of Hamiltonians 
\be
H^{(s)} = - \frac{d^2}{dx^2} + V^{(1)} (x, a_s)+
\sum_{k=1}^{s-1} R(a_k)
\ee
by repeated use of the shape invariance condition. Then, assuming that the ground state energy of the first Hamiltonian is zero and using the fact that spectra of the Hamiltonians 
$H^{(s)}$ and $H^{(s+1)}$ are the same, except of the
lowest energy of $H^{(s)}$, one gets \p{cont-sp}.

 \end{itemize}

To summarize, the technique of $N=2$ SQM
allows one to find a spectrum of a ``usual" nonsupersymmetric Hamiltonian $H^{(1)}$ by purely algebraic methods,
provided the corresponding potential satisfies
the shape invariance condition \p{SI-cont}.
Since the  discussion above 
was for the  case when the space coordinate $x$
is continuous, it is natural to consider
a modification of this method for the case
when SQM is defined on a lattice  \cite{Nicolai:1976xp}--\cite{deCrombrugghe:1982nmd}.
To this end, we use
the procedure of an ``exact discretization"
of the Schr\"odinger equation given in \cite{Tarasov-1}--\cite{Chou}
and find the factorization of the ``discrete"
 Hamiltonian 
in terms of the  of two ladder operators\footnote{For alternative methods
of factorization of second order differential equations, see \cite{Dobrogowska} and references therein.}.
This step actually means that we are considering
a discrete version of $N=2$ SQM.
Then we present a discretized version of 
the  shape invariance condition 
\p{SI-cont}
which allows one to find the  
wave functions and 
energy spectra of Hamiltonians, whose potentials obey this condition.
Finally,  we illustrate  this  technique
on the example of the Coulomb potential.
The last Section contains  a brief discussion of some open problems and possible generalizations of our results.

\section{Discretization} \label{disc} \setcounter{equation}0
We follow the procedure of the ``exact discretization", proposed by Tarasov \cite{Tarasov-1}--\cite{Tarasov-3}. 
The advantage of this discretization scheme is that an immediate correspondence 
between continuous differential equations and discrete difference equations 
can be achieved. Namely,  the main algebraic properties of differential operators,
like the semigroup property, Leibniz rule and differentiation of polynomials carry 
over to the difference operators. Furthermore, discretized solutions of differential
equations are solutions of the corresponding difference equations. All these properties 
are obtained without a limiting process, %$\lim_{a \rightarrow 0}$ 
in which the lattice constant  of the discrete model goes to zero. 

In the following, we summarize the main steps of
the ``exact discretization" \cite{Tarasov-1}--\cite{Tarasov-3}.

Given a wave function $\tilde{\psi}(x)$ one starts by computing its Fourier transform 
\begin{align}\label{eq:FT}
\mathcal{F}\left\lbrace \tilde{\psi}(x) \right\rbrace = \tilde{\psi}(k) =\frac{1}{2\pi}\int_{-\infty}^{\infty}dx \tilde{\psi}(x)e^{-ikx}
\end{align} 
and defines a new function 
\be \label{nf}
\psi(k)=\tilde{\psi}(k)\left(\Theta \left(k+\frac{\pi}{a}\right) - \Theta\left(k-\frac{\pi}{a}\right)\right),
\ee
where $a>0$ is a lattice constant and 
 $\Theta(k)$ is the Heaviside step function, $\Theta(k)=1$ for $k \geq 0$, and
$\Theta(k)=0$ for $k < 0$.  The new function  is equivalent to the former one $\psi(k)=\tilde{\psi}(k)$ within the interval $k\in [-\frac{\pi}{a}, +\frac{\pi}{a}]$. The Fourier transform of the function \p{nf} can be thought of as the coefficients of a discrete Fourier series %$\mathcal{F}_{a}$
\begin{align}\label{eq:FS}
\psi[n]=\frac{a}{2\pi}\int_{-\frac{\pi}{a}}^{\frac{\pi}{a}}dk \psi(k)e^{ik na} \equiv \mathcal{F}^{-1}_{a}\left\lbrace \psi(k) \right\rbrace.
\end{align}
Using the relation $\mathcal{F}\left\lbrace i\frac{\partial}{\partial x} \right\rbrace = k$ one can define a difference operator $\Delta^m$ which is a discrete analog of the $m_{\text{th}}$ derivative $D^m$
\begin{align}
\frac{\mathcal{F}^{-1}\left\lbrace \mathcal{F}_a \left\lbrace \Delta^m \right\rbrace \right\rbrace}{a^m}=D^m.
\end{align}
When acting on a ``trial" function $f[n] $ one gets
\begin{align}
\frac{\mathcal{F}^{-1}\left\lbrace \mathcal{F}_a \left\lbrace \Delta^m f[n] \right\rbrace \right\rbrace}{a^m} = D^m f[n] = \frac{\partial^m}{\partial x^m} \mathcal{F}^{-1}_a \left\lbrace \tilde{f} (k) \right\rbrace
\end{align}
and it follows that
\begin{align}\label{eq:convolution}
\mathcal{F}_a \left\lbrace \Delta^m f[n] \right\rbrace = (iak)^m \mathcal{F}_a \left\lbrace f[n] \right\rbrace.
\end{align}
This suggests that $\Delta^m f[n]$ has to be a discrete convolution
\begin{align}
\Delta^m f[n] = \sum^{\infty}_{j=-\infty}K_m[j]f[n-j] = \sum^{\infty}_{j=-\infty} K_m[j+n]f[n]
\end{align}
with $K_m$ being the kernel. Using \eqref{eq:convolution} one finds that the Fourier transform of the kernel should yield $\mathcal{F}_a\left\lbrace K_m \right\rbrace =(iak)^m $. 
Finally, the solutions of the 
equation \p{eq:convolution} for the
first- and for the second-order difference operators are
\bea \label{Delta-d}
&&\Delta^1 f[n] 
= \sum^{\infty}_{j=-\infty, j\neq 0} \frac{(-1)^j}{j}f[n-j], 
\label{Delta-d-1} \\
&&\Delta^2 f[n] = -\sum^{\infty}_{j=-\infty, j \neq 0} \frac{2(-1)^j}{j^2}f[n-j] -\frac{\pi^2}{3}f[n]. \label{Delta-d-2} 
\eea
One can show that these difference operators obey the semigroup property
\be \label{semigr}
\Delta^2 = \Delta^1 \Delta^1 
\ee
as well as the Leibniz rule
\be \label{leib}
\Delta^1(fg) = \Delta^1(f)g + f\Delta^1(g).
\ee 
In terms of the difference operator 
\p{Delta-d-2}
and the Fourier transforms  \eqref{eq:FS}, one finds a discretized Schr\"odinger equation 
\begin{align} \label{disc-S}
i\hbar \frac{d \psi[n](t)}{d t} = \frac{\hbar ^2}{m a^2} \sum^{\infty}_{j=-\infty, j\neq 0} \frac{2(-1)^j}{j^2}\psi[n-j](t) + {V}\psi[n] (t) \equiv H \psi[n] (t)
\end{align}
with
\be
V[n] = \frac{2\pi^2}{3a^2 } + \tilde{V}[n]
\ee
and $\tilde{V}[n]$ being 
a discretization of the potential which enters the continuous Schr\"odinger equation.

\section{ N=2 SQM on a lattice} \label{sqml} \setcounter{equation}0 \label{lsqm}
In order to 
construct a discretized version of $N=2$ SQM, we use the difference operator \p{Delta-d-1}
and define the ladder operators
\begin{align}
A_1[n] = \frac{\hbar}{\sqrt{2m}}\Delta^1 + W[n], \quad A_1[n]^\dagger =-\frac{\hbar}{\sqrt{2m}}\Delta^1 + W[n].
\end{align}
From these ladder operators we construct a lattice Hamiltonian
\begin{align} \label{discsusy}
H^{(1)}[n]=A^\dagger_1[n] A_1 [n] = -\frac{\hbar^2}{2m}\Delta^2 + V^{(1)}[n]
\end{align}
with the potential 
\begin{align}
V^{(1)}[n]=W^2[n] -\frac{\hbar}{\sqrt{2m}} \Delta^1 W[n].
\end{align}
Equation \p{discsusy} holds because the difference operator fulfills the semigroup property \p{semigr}
and the Leibniz rule \p{leib}.

The ground state wave function is given by a discrete analogue of the equation \p{exactsusy},
which yields
\begin{align}\label{eq:gswave}
W[n] \psi^{(1)}_0[n]  = -\frac{\hbar}{\sqrt{2m}} \Delta^1 \psi^{(1)}_0[n].
\end{align}  
As it was the case for the continuous $N=2$ SQM described in the Introduction,
one can construct a hierarchy of Hamiltonians
with the same properties of the energy levels and  wavefunctions as in
\p{en-lev-cont}. In particular,
the second Hamiltonian in the hierarchy is
\begin{align}
H^{(2)}[n]=A_1[n] A^\dagger_1[n] = -\frac{\hbar^2}{2m}\Delta^2 + V^{(2)}[n]
\end{align}
with the potential
\begin{align}
V^{(2)}[n]=W^2[n] +\frac{\hbar}{\sqrt{2m}} \Delta^1 W[n]
\end{align}
and so on.
Finally, in order to find the spectrum of the
original Hamiltonian $H^{(1)}[n]$ we introduce
a condition of the shape invariance, 
\begin{align}\label{SI-disc}
W^2[n](a_1) + \Delta^{1}W[n](a_1)  = W^2[n](a_2) - \Delta^{1}W[n](a_2) + R(a_1),
\end{align}
where $\phi$ is a function of the parameters $a_1$ and $a_2$ which are present in the Hamiltonians
$H^{(1)}$ and $H^{(2)}$.

\section{Solving the discrete Coulomb potential} \setcounter{equation}0 \label{cp}

As it follows from the discussion given in  Sections \ref{disc}
and \ref{sqml}, one can immediately apply the technique
of discretized $N=2$ SQM for finding
of energy spectra of the one-dimensional Quantum Mechanical models,
whose  list (in case of continuous space coordinate $x$) is
given in \cite{Dutt:1986va}.
Indeed, the crucial requirement for  applying of the technique 
of $N=2$ SQM for finding of the energy spectra
is that  corresponding potential should satisfy
the shape invariance condition 
\p{SI-cont}. On the other hand, the procedure of the ``exact discretization", 
essentially maps this equation
into its discrete counterpart \p{SI-disc}.
Therefore, the potentials given in \cite{Dutt:1986va}
will satisfy the discretized shape invariance condition after
Taylor expanding and replacing 
the continuous coordinate with the lattice constant times an integer  $n$.
Finally, the properties \p{semigr} and \p{leib}
of the difference operators provide a possibility of 
the corresponding factorization and for a construction
of the hierarchy of Hamiltonians which leads to the 
expression for the energy spectrum given in \p{cont-sp}.

As an illustration of the discussion above, let us consider as an example the discretization of the Coulomb potential in some detail.
 The Coulomb potential 
 is an example of a Quantum Mechanical system  which in the continuous case
has been solved  using the technique of $N=2$ SQM \cite{Valence:1990}.
Here we shall present a similar discussion for the energy spectrum of the corresponding 
discretized model. 
Recall that by exploiting spherical symmetry, the Coulomb problem becomes effectively one-dimensional and the potential can be written in natural units as
\be
V(r) = -\frac{1}{r} + \frac{\ell(\ell+1)}{2 r^2}, \quad r>0.
\ee
In the discrete case, we replace the continuous radius by an integer $r \rightarrow a n$. For simplicity we shall
set the lattice constant to unity $a=1$. This potential can be factorized by the superpotential
\be
W[n]= \frac{1}{\ell +1} - \frac{\ell +1}{n}.
\ee
Using the power law property of the first difference 
operator\footnote{
The proof for differences of a power law function, $\Delta^1 n^{k} = k n^{k-1}$, given in \cite{Tarasov-2} holds only for integer $k \in \mathbb{N}$. However, one can easily extend it for $k\in \mathbb{R}$ using generalized binomial coefficients. The reason that it works stems from the fact that only a single term in the summation over the binomial coefficients remains finite.}
$\Delta^1 n^{k} = k n^{k-1}$  and the fact that
for a constant $c$ one has
$\Delta^1 c = 0$,  we  construct  two partner potentials
\begin{align}
V^{(1)}[n]=&\frac{1}{(\ell+1)^2} +\frac{(\ell+2)(\ell+1)}{n^2} - \frac{2}{n} \\
V^{(2)}[n]=&\frac{1}{(\ell+1)^2} +\frac{\ell(\ell+1)}{n^2} - \frac{2}{n} .
\end{align}
These two potentials are indeed related by the  shape invariance
condition \eqref{SI-disc}. The role of the parameter $a_1$
is played by $\ell$,
while the function $\phi$ which relates the 
parameters $a_2$ and $a_1$ in the partner Hamiltonians is given by
$a_2 = \phi(\ell)=\ell +1$. 
We can further identify the ground state energy $E^{(1)}_0=\frac{1}{(\ell+1)^2}$ and the "rest" function as
\be
R(\ell)=\frac{2\ell+3}{(\ell+1)^2(\ell+2)^2}.
\ee
According to \p{cont-sp} the full spectrum can then be found
\be
E^{(1)}_n = E^{(1)}_0+\sum_{\ell=1}^{n}R(\ell)= \frac{1}{(\ell+1)^2}+
\sum_{m=1}^n\frac{2(\ell+m)+1}{(\ell+m)^2(\ell+m+1)^2}.
\ee
The ground state wave function can also be easily obtained in the discrete case by 
by using the equation \eqref{eq:gswave}. 
Making an ansatz for the ground state wave function 
\be 
\psi^{\ell=0}_0[n]=\sum_{j=0}^\infty c_j n^j
\ee
one finds 
the condition for the unknown coefficients
\be
\frac{c_0}{n}+\sum_{j=0}^{\infty} \left(jc_{j+1}+c_j \right)n^j = 0.
\ee
This equation means that $c_0=0$, while the
other  coefficients obey recurrence relations
\be
jc_{j+1}+c_j = 0,
\ee
which can be solved by
\be
c_j = \frac{(-1)^{j-1}}{j!}N
\ee
with $N$ some undetermined constant. We finally find the ground state wave function
\be
\psi^{\ell=0}_0[n]= N n e^{-n}
\ee
which represents a discrete analog of the well known result of the continuum case.

\section{Conclusions}
In this  paper, we 
considered 
the problem of finding of the energy spectrum 
for lattice Hamiltonians
in terms a discretized version 
of $N=2$ Supersymmetric Quantum Mechanics, by applying of the corresponding shape invariance condition.
We restricted ourselves with the consideration of the simplest case of the one-dimensional
Quantum Mechanical models.
It would be very interesting to generalize this technique
to the case of two- and three-dimensional systems,
 appropriately  generalizing the procedure
 of ``exact discretization" of the Schr\"odinger equation
and considering a discretization of the shape invariance condition
for higher-dimensional $N=2$ SQM
  \cite{Andrianov:2012vk}.
  This generalization could allow an application
  of the suggested technique for finding of energy spectra of some realistic physical examples.
  
Another interesting generalization of  our results
could be  a construction and further
study  of discrete versions of
 systems with higher number of supersymmetries.
 An example of such systems are one dimensional \cite{Ivanov:1990cz} and multidimensional \cite{Donets:1999jx}
 $N=4$ 
 Supersymmetric Quantum Mechanics. Let us note, that 
 in these systems the 
  supersymmetry algebra can also contain  nonzero central charges, which gives an interesting
  structure for vacuum ground states (so -- called partial supersymmetry breaking)\footnote{A discussion of isospectral Hamiltonians in the framework   $N=4$ SQM can be found
 in \cite{Berezovoi:1991xc}. An application to particular
 $N=4$ SQM models is given in \cite{Donets:1999jx}--\cite{Ivanov:2020bji}.}. 
 An application of our approach to  models with nonlinear supersymmetry and hidden suppersymmatry
 (considered in the continuous case in \cite{Correa:2008hc}--\cite{Correa:2006je})
 is another interesting problem to address.

\vskip 0.5cm

\noindent {\bf Acknowledgments.} 
We are grateful to Yasha Neiman for useful  discussions and comments on the manuscript.
 J.S. received financial support from the Theory of Quantum Matter 
Unit of the Okinawa Institute of Science and Technology Graduate University
(OIST).
The work of M.T.  was supported by the Quantum Gravity Unit
of the Okinawa Institute of Science and Technology Graduate University
(OIST).

\providecommand{\href}[2]{#2}\begingroup\raggedright
\endgroup

\end{document}